\documentclass[11pt]{article}
\fontfamily{times}
\usepackage{graphicx}
\usepackage{geometry}
\usepackage{xcolor}
\definecolor{gray}{rgb}{.6,.6,.6}
\definecolor{orange}{rgb}{1,0.5,0}
\definecolor{grayish}{rgb}{.975, .975, .975}
\definecolor{dkgray}{rgb}{.375, .375, .375}
\definecolor{dkgreen}{rgb}{0,0.6,0}
\definecolor{mauve}{rgb}{0.58,0,0.82}
\definecolor{dkblue}{rgb}{0, 0, .5}
\usepackage{xspace}
\usepackage{hyperref}
\usepackage{natbib}

\usepackage{listings}
\let\proglang=\textsf
\newcommand{\pkg}[1]{{\fontseries{b}\selectfont #1}}

\newcommand{\R}{\proglang{R}\xspace}

\begin{document}
\quad
\small{
\begin{center}
  \textbf{Parallel Statistical Computing with R: \\
    An Illustration on Two Architectures}\footnote{This work used
    resources of the Oak Ridge Leadership Computing Facility at the
    Oak Ridge National Laboratory, which is supported by the Office of
    Science of the U.S. Department of Energy under Contract
    No. DE-AC05-00OR22725. This material is based in part upon work
    supported by the National Science Foundation Division of
    Mathematical Sciences under Grant No. 1418195.}
\end{center}

\begin{center}
{George Ostrouchov*}\\
{Oak Ridge National Laboratory and University of Tennessee}\\
\vspace{0.5cm}
{Wei-Chen Chen}\\
{pbdR Core Team}\\
\vspace{0.5cm}
{Drew Schmidt}\\
{Oak Ridge National Laboratory}\\
\end{center}

\begin{center}
{\bf Abstract}
\end{center}

\vspace{-1em}
\noindent To harness the full benefit of new computing platforms, it
is necessary to develop software with parallel computing
capabilities. This is no less true for statisticians than for astrophysicists.
The R programming language, which is perhaps the most popular software
environment for statisticians today, has many packages available for parallel
computing.  Their diversity in approach can be difficult to navigate.
Some have attempted to alleviate this problem by designing common interfaces.
However, these approaches offer limited flexibility to the user;
additionally, they often serve as poor abstractions to the reality of modern
hardware, leading to poor performance.
We give a short introduction to two basic
parallel computing approaches that closely align with hardware
reality, allow the user to understand its performance, and provide
sufficient capability to fully utilize multicore and multinode
environments.

We illustrate both approaches by working through a simple example
fitting a random forest model. Beginning with a serial algorithm, we
derive two parallel versions. Our objective is to illustrate the use
of multiple cores on a single processor and the use of multiple
processors in a cluster computer. We discuss the differences between
the two versions and how
the underlying hardware is used in each case.\\

}\\

\noindent {\bf 1. Introduction}

\noindent For over a decade, processor speeds have been nearly constant.
Instead of chasing clock speeds as in the 1990's, hardware manufacturers have
taken to adding more cores per processor in their new product lines.  Like it
or not, this makes parallel computing a necessary skill for developers if they
wish to take full advantage of new hardware. This trend and the availability
of ever larger data sets are fueling the development of many \R packages
dealing with some form of parallelism.

On the other hand, it has been more than 30 years since
the first ``production'' multiprocessors by Intel and nCube were
delivered. Soon after, a large community has formed around practical
research in parallel numerical
computing~\citep{heath1987hypercube}, which produced many
publications by the late 1980's (see
bibliography~\cite{OrtegaEtAl1989}) including some in statistical
computing~\citep{Ostrouchov87a,EddySchervish87,Schervish88}.
While the activity in statistical computing did not continue at the
same pace over the next two decades, work in numerical linear algebra
and in the solution of partial differential equations produced an
impressive collection of scalable\footnote{ {\em Scalability} is a
  software's ability to take advantage of more parallel resources as
  they become available.}  numerical
libraries (BLACS, PBLAS, ScaLAPACK, Trilinos, MAGMA, DPLASMA, PETSc, etc.).
Some of these in turn lead to the production of vendor libraries, such
as Intel MKL, AMD ACML, and Cray LibSci, tuned for specific hardware.

There are now over 50 packages on CRAN that address some form of
parallelism. Diverse concepts are introduced,
resulting in a somewhat bewildering collection that makes for
difficult decisions on the approach to take. While some expose the
underlying hardware
to the user, others go to great lengths to mask the reality in favor
of a common interface. When the user is not aware of what goes on
``behind the scenes'',
an application can slow down rather than speed up, even as
more resources are used.

In this short tutorial, we do not attempt to classify all the
approaches and refer the readers to other more comprehensive
reviews~\citep{Schmidberger2009,Wang2015,Matloff2015}. The reviews are
useful but they still present too many options that a new user will find
difficult to navigate. Here, we choose two basic concepts that have a
long history in parallel computing and a modern interface in \R. We
demonstrate their use by describing how to parallelize a random forest
code implementation.

The two basic concepts are (1) the use of a single processor with
multiple cores via the \pkg{parallel} package, and (2) the use of
multiple processors through a batch submission to a cluster computer
using MPI via the \pkg{pbdMPI} package. It is possible to combine
these two approaches where a multi-node run will use multiple cores within each
node, if care is taken not to oversubscribe the cores. Mastering these two
concepts provides a basis from which the \R
user can graduate to more complex parallel concepts such as the use of
scalable libraries and advanced profiling capablities of the pbdR
project~\citep{pbdR2012}, which can take the user all the way to
world's largest parallel systems~\citep{Schmidt20171} all within the
comfort of \R.
\\

\noindent {\bf 2. A Serial Random Forest Example}

\noindent Random forest is a popular classification technique
that is still among the best today~\citep{Breiman2001}.
It is a type of ensemble learning that can be used for
classification and regression. Given a response variable
and predictors, simple models are trained on random
subsets of the predictors providing some control of
overfitting. Predictions are made by averaging the simple
models.

\begin{figure}[tb]
  \begin{center}
    \begin{minipage}{0.4\textwidth}
      \includegraphics[width=0.6\textwidth]{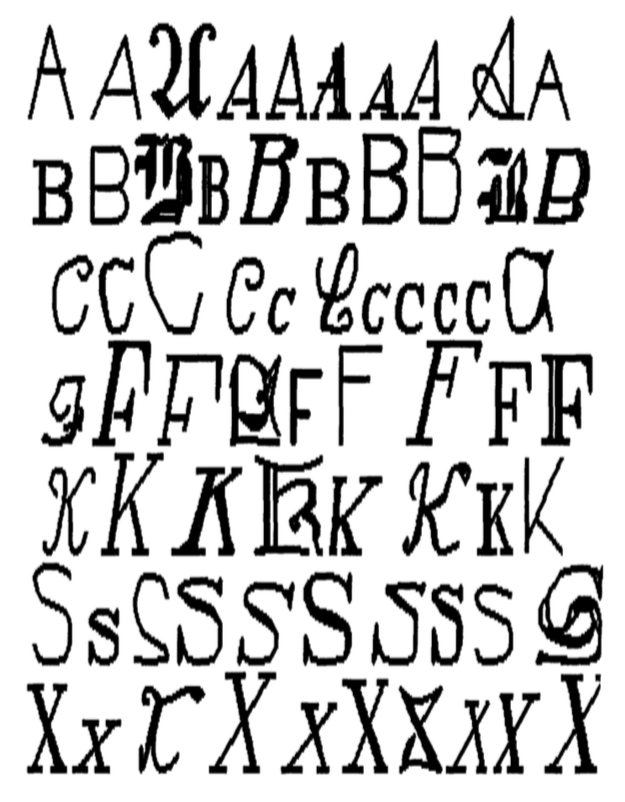}
      \vspace{1em}
    \end{minipage}
    \begin{minipage}{0.5\textwidth}\scriptsize
      \begin{verbatim}
[,1]	lettr	capital letter
[,2]	x.box	horizontal position of box
[,3]	y.box	vertical position of box
[,4]	width	width of box
[,5]	high	height of box
[,6]	onpix	total number of on pixels
[,7]	x.bar	mean x of on pixels in box
[,8]	y.bar	mean y of on pixels in box
[,9]	x2bar	mean x variance
[,10]	y2bar	mean y variance
[,11]	xybar	mean x y correlation
[,12]	x2ybr	mean of x^2 y
[,13]	xy2br	mean of x y^2
[,14]	x.ege	mean edge count left to right
[,15]	xegvy	correlation of x.ege with y
[,16]	y.ege	mean edge count bottom to top
[,17]	yegvx	correlation of y.ege with x
      \end{verbatim}
    \end{minipage}
  \end{center}
  \vspace{-2em}
  \caption{Letter Recognition data (image: \citep{Frey1991}, description: \pkg{mlbench} package).}
  \label{fig:LR}
\end{figure}

We use the famous Letter Recognition data set~\citep{Frey1991},
which is available in the \R package \pkg{mlbench}~\citep{Leisch2010}.
It contains 16 measurements on 20,000
corrupted capital letters of the alphabet indicated in column 1 of the
resulting 20,000 $\times$ 17 matrix (see Fig.~\ref{fig:LR}).

The CRAN package \pkg{randomForest} \citep{Liaw2002}
contains the necessary functions to easily implement a serial
training and prediction algorithm
that we show in Fig.~\ref{fig:rf-serial}.
\begin{figure}[tb]
\begin{lstlisting}
library(randomForest)
library(mlbench)
data(LetterRecognition)
set.seed(seed = 123)

n <- nrow(LetterRecognition)
n_test <- floor(0.2 * n)
i_test <- sample.int(n, n_test)
train <- LetterRecognition[-i_test,]
test <- LetterRecognition[i_test,]

rf.all <- randomForest(lettr ~ ., train, ntree = 500, norm.votes = FALSE)
pred <- predict(rf.all, test)
cat("Proportion Correct:", sum(pred == test$lettr)/n_test, "\n")
\end{lstlisting}
  \vspace{-1ex}
  \caption{Serial random forest algorithm using functions
    from \pkg{randomForest}.}
  \label{fig:rf-serial}
\end{figure}
We train the random forest on 80\% of the data and test it on the
remaining 20\%.

We set the random number generator seed in line 4 after
 loading the necessary packages and the
data. Lines 6-8 select a random 20\% of the
matrix rows for testing. Then, lines 9 and 10 subset the training and
the testing
data sets. Line 12 trains the forest and line
13 makes the letter predictions for the test subset. Finally line 14
prints the proportion of correct predictions. Simple.

Both parallel approaches
split training by building blocks of trees in parallel and use the
function {\tt combine()} to build one forest from all the
blocks. Then, prediction with the combined forest is done on blocks of
testing data rows in parallel. That is, we split the models for
training and we split the data for testing. This means that all
processors must see all the training data but the testing data is
split among the processors.
\\

\noindent {\bf 3. Using Multiple Cores  with \pkg{parallel} Package}

\noindent We find {\tt mclapply()} to be the most useful among the
methods in the \pkg{parallel} package on unix and mac platforms. It is
not available for Windows. To use it (see Fig~\ref{fig:rf-mc}),
\begin{figure}[bt]
\begin{lstlisting}
library(randomForest)
library(mlbench)
data(LetterRecognition)
library(parallel)
RNGkind("L'Ecuyer-CMRG")  # mc
set.seed(seed = 123)

n <- nrow(LetterRecognition)
n_test <- floor(0.2 * n)
i_test <- sample.int(n, n_test)
train <- LetterRecognition[-i_test, ]
test <- LetterRecognition[i_test, ]

nc <- detectCores()  # mc
ntree <- lapply(splitIndices(500, nc), length)  # mc
rf <- function(x) randomForest(lettr~., train, ntree=x, norm.votes=FALSE)# mc
rf.out <- mclapply(ntree, rf, mc.cores = nc)  # mc
rf.all <- do.call(combine, rf.out)  # mc

crows <- splitIndices(nrow(test), nc)  # mc
rfp <- function(x) as.vector(predict(rf.all, test[x, ]))  # mc
cpred <- mclapply(crows, rfp, mc.cores = nc)  # mc
pred <- do.call(c, cpred)  # mc
cat("Proportion Correct:", sum(pred == test$lettr)/n_test, "\n")
\end{lstlisting}
  \vspace{-1ex}
  \caption{Parallel random forest algorithm using functions
    from \pkg{randomForest} and \pkg{parallel}.}
  \label{fig:rf-mc}
\end{figure}
we make a function out
of the code that needs to run in parallel ({\tt rf} and {\tt rfp}) and
construct a vector of
parameters that informs the parallel instances ({\tt ntree} and {\tt
  crows}). The unix system {\tt fork}
is used by {\tt mclapply()} to spawn parallel child processes. Each
process can access
the same data as the parent process in a
{\it copy-on-write} fashion. This means the function input data is not
copied unless it is modified. This is particularly convenient in
statistical computing as the data is typically not
modified. Consequently the data is shared in memory, not
duplicated, and its processing is done in parallel. This is how a
multicore shared memory processor is intended to be used.

As in the serial code, we load the packages and the data in the first four
lines, then we select a parallel-capable random number
generator and set its seed in lines 5 and 6. \R's default random
number generator might work most of the time but there is no guarantee
that the parallel streams will not overlap even if different seeds are
used.

Lines 8 to 12 select the testing and training data sets as the serial
code. We detect the number of cores present on the processor in 14 and
split the 500 into that many nearly equal pieces. This vector ({\tt
  ntree}) tells
each core how many trees to build in line 17.
The result of {\tt mclapply()} is a list with a forest block from each
core used. The function {\tt combine} then puts them all together into
one predictor forest in 18.

For prediction on the test data matrix, we split its row indices in line 20
into as many groups as there are cores. The prediction is done in
parallel in line 22. Line 23 concatenates resulting list of prediction
groups and the proportion correct is printed in line 24.

Note that in both forest generation and prediction we made a
few large chunks instead of many small chunks to reduce function calls
and results collection overhead in {\tt mclapply()}.
\\

\noindent {\bf 4. Using Multiple Processors with \pkg{pbdMPI} Package}

\noindent Multiple nodes in a cluster architecture traditionally
force the programmer to consider how a problem and its data are
partitioned across the machine. MPI \citep{mpi3}
is the standard for communicating data between processors.   While
managing this seems like a daunting challenge, the HPC community has
found a way to deal with it by writing a single program that runs
asynchronously on each processor. This is called single program,
multiple data (SPMD) programming. Each SPMD instance here is a
separate \R process. It is a natural extension
of serial programming. With some experience, SPMD is easily seen as
the best way to write complex parallel programs on distributed
architectures. The \pkg{pbdMPI} package (\cite{Chen2012pbdMPIpackage})
is a simplified high-level interface to MPI.

We say {\it processors} because MPI works between {\it nodes} or
between {\it cores}, making SPMD programming universal to shared and
distributed memory. But {\it fork} works only in shared memory so why
would anyone bother with {\tt mclapply()}?  The answer is memory
because a forked process uses data as {\it copy-on-write}, avoiding
the possible data duplication incurred when several cores on the same
node need the same data. There are also differences in overhead but
the application, hardware, and system software ultimately determine
which is faster.

The use of MapReduce in Spark and Hadoop distributed systems is
popular in cloud computing but comparisons on cluster computers with
multicore processors show that SPMD approaches, exemplified by the pbdR
project, provide much faster, more extensible, and more scalable
solutions \citep{Schmidt20171,Xenopoulos2016}.

To produce the parallel SPMD code in Fig~\ref{fig:rf-mpi}
\begin{figure}[tb]
\begin{lstlisting}
library(randomForest)
library(mlbench)
data(LetterRecognition)
library(pbdMPI)
comm.set.seed(seed = 123, diff = FALSE)

n <- nrow(LetterRecognition)
n_test <- floor(0.2 * n)
i_test <- sample.int(n, n_test)
train <- LetterRecognition[-i_test, ]
test <- LetterRecognition[i_test, ][get.jid(n_test), ]

comm.set.seed(seed = 1e6 * runif(1), diff = TRUE)
my.rf <- randomForest(lettr ~ ., train, ntree = 500 %/% comm.size(), norm.votes = FALSE)
rf.all <- do.call(combine, allgather(my.rf))
pred <- as.vector(predict(rf.all, test))

correct <- reduce(sum(pred == test$lettr))
comm.cat("Proportion Correct:", correct / (n_test), "\n")
finalize()
\end{lstlisting}
  \vspace{-1ex}
  \caption{Parallel random forest algorithm using functions
    from \pkg{randomForest} and \pkg{pbdMPI}.}
  \label{fig:rf-mpi}
\end{figure}
we generalize the serial code so that it can cooperate with several
copies of itself running in parallel. When running $n$ copies of the
code, each copy has a different {\it rank} number (0 to $n-1$), which
is returned by the function {\tt comm.rank()} and is used to
distinguish what data is operated on or read by a particular
copy. However, high level communication operations are available
in \pkg{pbdMPI} so that the rank wrangling is done completely
benind the scenes without the user involvement, as it is in our
example here. To run the code in Fig.~\ref{fig:rf-mpi} on 4
processors, put it in a file named {\tt rf-mpi.r} and use the
following command at the shell prompt
\begin{lstlisting}[language=sh,aboveskip=2mm,belowskip=1mm,backgroundcolor=\color{white}]
mpirun -np 4 Rscript rf-mpi.r
\end{lstlisting}
On a cluster with a PBS job scheduler, this works after an interactive
{\tt qsub} allocation or in a shell script for a PBS
submission. You may need to consult with your cluster administrator as
there are different job schedulers. On a desktop of a laptop this
works at the shell prompt after installation of \pkg{pbdMPI}.

Every line in Fig.~\ref{fig:rf-mpi} runs on every
processor, sometimes duplicating what others are doing and sometimes
working on different data. In line 5, all set the same seed after loading
the libraries. Loading \pkg{pbdMPI}  makes the {\tt
  comm.set.seed()} parallel random number generator function
available. We first set {\tt diff~=~FALSE}
because we want everyone to have the same training
data, which is separated in lines 7 to 11. Line 11 does an additional
subsetting of the test data so that
each rank gets a different set of rows. The function {\tt get.jid(n)}
returns the indices for one rank out of $n$ total indices, using
rank information behind the scenes.

Line 13 sets a new seed with {\tt diff~=~TRUE} so that line 14 builds
a different set of trees on each rank. The function {\tt comm.size()}
returns the total number of ranks to divide the total of 500 trees
between the ranks. Line 15 combines the trees generated by each rank
to all ranks, first as a list returned by {\tt allgather()} and then
as a forest put together by the {\tt combine()} function. Line 16 does
the prediction using the combined forest on this rank's rows of the
test data.

Line 18 computes the local number of correct predictions and
the {\tt reduce()} function adds across the ranks, sending the result
to rank 0 (the default). An {\tt allreduce()} would send the result
to all ranks if further parallel processing was needed. Finally, {\tt
  comm.cat()} writes the result from rank 0 (the default) and {\tt
  finalize()} properly exits from {\tt mpirun}.\\

\noindent {\bf 5. Benchmarks and Conclusions}

\begin{figure}[tb]
  \begin{center}
    \includegraphics[width=.55\textwidth]{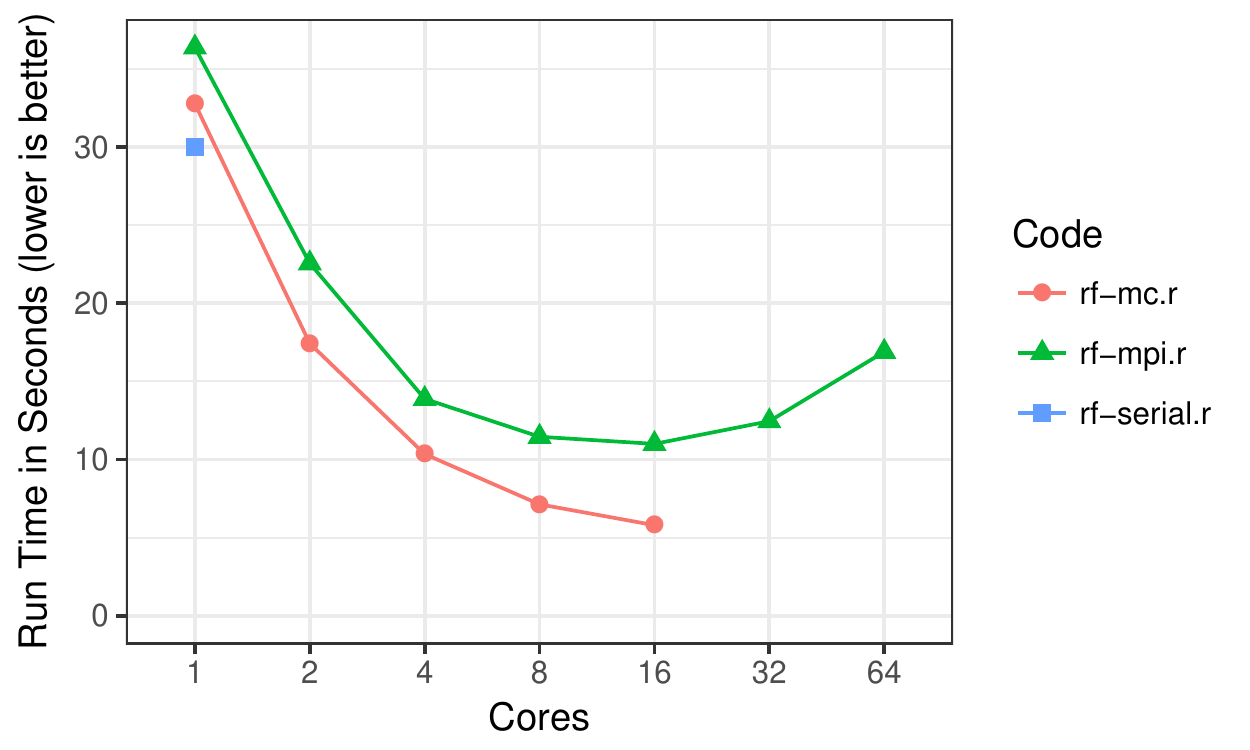}
    \vspace{-2ex}
    \caption{Random forest benchmark on the Oak Ridge
      National Laboratory machine Rhea.}
    \label{fig:results}
  \end{center}
\end{figure}

\noindent We examine the performance of the implementations on the Oak
Ridge National Laboratory cluster Rhea. Rhea is a 512-node
commodity-type Linux cluster. Each node contains two 8-core 2.0 GHz
Intel Xeon processors and 128 GB of memory.
Figure~\ref{fig:results} shows the results of the benchmark.

We see that for one core, the serial
implementation is a bit faster than both parallel
implementations. This is not too
surprising, since one can readily see by comparing
Figures~\ref{fig:rf-serial}, \ref{fig:rf-mc},
and~\ref{fig:rf-mpi} that the parallel versions have additional computational
overhead.  However, as we increase the number of cores, the run times begin
to drop proportionally, up until about 16 cores.  Perhaps the ``best'' run,
in the sense of performance improvement per core added, is achieved at 8
cores for both parallel codes.  But certainly the performance improves up to 16 cores.  The problem
is fairly small for the larger core counts.  Essentially, the parallel
versions begin to run out of work at about 8 cores,
but a little more performance can be squeezed out of 16 cores.

The {\tt mclapply()} parallel version performs slightly better than
the MPI version throughout, likely due to somewhat higher overhead
involved in MPI communications between \R instances compared to
forking an \R process. This comparison on Rhea would likely persist
with a larger problem that does not run out of work at 16 cores,
however the MPI version would be able to use more nodes than {\tt
  mclapply()} and continue to scale to eventually produce lower run
times. Optimally, one would write a third parallel version that would
combine the two approaches, giving optimal scaling within nodes with
{\tt mclapply} together with the ability to solve larger problems that
need the combined memory of multiple nodes with \pkg{pbdMPI}
package. We leave this challenge to the reader as the next step before
considering more complex algorithms that rely on scalable matrix
computations supported by the pbdR project \citep{pbdR2012}.


\bibliographystyle{abbrvnat}
\bibliography{programming}

\end{document}